\documentclass[prc,twocolumn,showpacs,preprintnumbers,amsmath,amssymb,superscriptaddress,floatfix,nofootinbib]{revtex4}
\usepackage{graphicx}
\usepackage{epsfig}
\usepackage{epstopdf}
\usepackage{hyperref}
\usepackage{amsmath}
\usepackage{amsfonts}
\usepackage{amssymb}%
\newcommand{\Slash}[1]{\ooalign{\hfil/\hfil\crcr$#1$}}

\begin{document}

\title{ Nucleon resonances in the $\pi^- p \to K^0 \Lambda$ reaction near threshold}

\author{Cheng-Zu Wu}

\affiliation{Department of Physics, Dalian University of Technology,
Dalian 116024, China} \affiliation{Institute of Modern Physics,
Chinese Academy of Sciences, Lanzhou 730000, China}

\author{Qi-Fang L\"u}

\affiliation{Department of Physics, Zhengzhou University, Zhengzhou, Henan 450001, China}

\author{Ju-Jun Xie} \email{xiejujun@impcas.ac.cn}

\affiliation{Institute of Modern Physics, Chinese Academy of
Sciences, Lanzhou 730000, China} \affiliation{State Key Laboratory
of Theoretical Physics, Institute of Theoretical Physics, Chinese
Academy of Sciences, Beijing 100190, China}

\author{Xu-Rong Chen}
\affiliation{Institute of Modern Physics, Chinese Academy of Sciences, Lanzhou 730000, China}

\begin{abstract}

We investigate the two-body reaction $\pi^- p \to K^{0} \Lambda$
within the effective Lagrangian approach and the isobar model. In
addition to the ``background" contributions from $t$-channel $K^*$
exchange, $u$-channel $\Sigma(1192)$ and $\Sigma^*(1385)$ exchanges,
and $s$-channel nucleon pole terms, the contributions from the
nucleon resonances $N^*(1535)$, $N^*(1650)$, and $N^*(1720)$ are
investigated. It is shown that the inclusion of these nucleon
resonances contributions leads to a good description of the
experimental total and differential cross sections data at low
energy region. The $s$-channel $N^*(1535)$, $N^*(1650)$, and
$N^*(1720)$ resonances and the $t$-channel $K^*$ exchange give the
dominant contributions below $W = 1.76$ GeV, while the $s$-channel
nucleon pole, $u$-channel $\Sigma(1192)$ and $\Sigma^*(1385)$
exchanges give the minor contributions.

\end{abstract}


\maketitle

\section{Introduction}

The study of the spectrum of the nucleon resonances and the
resonances coupling constants from the available experimental data
are two of the most important issues in hadronic physics and they
are attracting much attention~\cite{Klempt:2009pi}. In the classical
quark models~\cite{Isgur:1978xj,Capstick:1986bm,Loring:2001kx}, a
rich spectrum of excited nucleon states is predicted. Many of these
nucleon resonances could be identified in $\pi N$ scattering.
However, there are still some of them have not been so far observed.

The strangeness production reaction $\pi^- p \to K^{0} \Lambda$ is a
good platform to study the properties of the nucleon resonances,
especially for those which have significant couplings to $\pi N$ and
$K \Lambda$ channels, because the $\pi^- p \to K^0 \Lambda$ reaction
is a pure isospin $I = 1/2$ two-body reaction channel in
meson-nucleon dynamics, and there are no contributions from the
isospin $I = 3/2$ $\Delta(1232)$ baryons. Hence, lots of
experimental data have been
accumulated~\cite{Baker:1978qm,Knasel:1975rr,Bertanza:1962pt,Baldini:1988ti},
where the total and differential cross sections of $\pi^- p \to K^0
\Lambda$ reaction are measured.

In response to this wealth of data, the theoretical activity has run
in parallel. In Ref.~\cite{Sibirtsev:2005mv}, the two-body reaction
$\pi^- p \to K^{0} \Lambda$ is investigated using the partial wave
amplitudes which are constructed from the available experimental
data. The total and differential cross sections of $\pi^- p \to
K^{0} \Lambda$ reaction can be well described by using these
amplitudes, which are essential for obtaining spin-flip and spin
non-flip amplitudes~\cite{Sotona:1988fm}. But, from these
amplitudes, it is difficult to get clear properties of some nucleon
resonances, because the partial wave amplitudes could have
contributions from nucleon resonances with different spins, and
individual contributions are more difficult to pin down. This
deficiency is also shown in Ref.~\cite{Ronchen:2012eg} where the
reactions $\pi N \to \pi N, \eta N, K \Lambda, K \Sigma$ are studied
simultaneously within an analytic, unitary, coupled-channel
approach. Moreover, it is pointed out that there are ambiguities of
the $\pi^- p \to K^0 \Lambda$ scattering amplitudes obtained from
the partial wave analysis, when only the observables of the
differential cross section and polarization are
measured~\cite{Anisovich:2013tij}.

The role played by the nucleon resonance $N^*(1535)$, which has
proved to be a controversial resonance for many years, in the $K
\Lambda$ production is crucial. The $N^*(1535)$ couples strongly to
the $\eta N$ channel~\cite{pdg2014} but a large $N^*(1535) K
\Lambda$ coupling has also been
deduced~\cite{GarciaRecio:2003ks,Liu:2005pm,Liu:2006ym,Geng:2008cv}
through the analysis of BES data on $J/\psi \to \bar{p}K^+\Lambda$
decay~\cite{Yang:2005ej} and COSY data on the $pp \to p\Lambda K^+$
reaction near threshold~\cite{Kowina:2004kr}. In
Refs.~\cite{Penner:2002ma,Penner:2002md,Shklyar:2005xg,JuliaDiaz:2006is},
the analyses of recent SAPHIR~\cite{Tran:1998qw,Glander:2003jw} and
CLAS~\cite{Nasseripour:2008aa} $\gamma p \to K^+ \Lambda$ data also
indicate a large coupling of the $N^*(1535)$ resonance to the $K
\Lambda$ channel. Furthermore, in a chiral unitary coupled channel
model it is found that the $N^*(1535)$ resonance is dynamically
generated, with its mass, width and branching ratios in fair
agreement with
experiment~\cite{GarciaRecio:2003ks,Kaiser:1996js,Inoue:2001ip,Nieves:2001wt,Doring:2009uc}.
This approach shows that the couplings of the $N^*(1535)$ resonance
to $K\Lambda$ channel could be large compared to those for $\pi N$.
We wish to argue in  this work that the $N^*(1535)$ resonance might
play a much wide role in associated strangeness production of $\pi^-
p \to K^0 \Lambda$ reaction.

In the present work, we study the two-body reaction $\pi^- p \to K^0
\Lambda$ within an effective Lagrangian approach and the isobar
model, which is an important theoretical method for describing
various processes in the resonances production
region~\cite{Tsushima:1996tv,Tsushima:1998jz,Shyam:1999nm,Xie:2007qt,Dai:2011yr,Liu:2011sw,Liu:2012ge,Lu:2013jva,Lu:2014rla,Xie:2014kja}.
In addition to the background contributions from $t$-channel $K^*$
exchange, $u$-channel $\Sigma(1192)$ and $\Sigma^*(1385)$ exchange,
and $s$-channel nucleon pole terms, we also investigate the
contributions from nucleon resonances $N^*(1535)$, $N^*(1650)$, and
$N^*(1720)$, which have significant couplings to $\pi N$ and $K
\Lambda$ channels~\cite{pdg2014}.

This article is organized as follows. In Sect.~II we present the
formalism and ingredients required for the calculation. The
numerical results and discussions are given in Sect.~III. A short
summary is
given in the last section.\\

\section{FORMALISM AND INGREDIENTS}

In this section, we introduce the theoretical formalism and
ingredients for calculating the $\pi^- p \to K^{0}\Lambda$ reaction
by using the effective Lagrangian approach and the isobar model. In
the following equations, we use $N^*_1$, $N^*_2$, and $N^*_3$, which
denote the $N^*(1535)$, $N^*(1650)$, and $N^*(1720)$, respectively.

\begin{figure}[htbp]
\begin{center}

\includegraphics[scale=0.5]{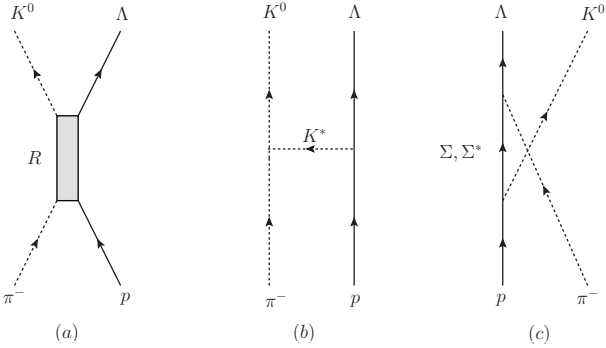}
\caption{Feynman diagrams for the reaction $\pi^- p \to
K^{0}\Lambda$. The contributions from $s$-channel nucleon resonance
$t$-channel $K^*$ exchange, and $u$-channel $\Sigma(1192)$ and
$\Sigma^*(1385)$ exchange are considered.} \label{Fig:feydg}
\end{center}
\end{figure}

The basic tree level Feynman diagrams for the $\pi^- p \to K^0
\Lambda$ reaction are depicted in Fig.~\ref{Fig:feydg}. These
include $s$-channel nucleon pole and nucleon resonances process
[Fig.~\ref{Fig:feydg} (a)], $t$-channel $K^*$ exchange
[Fig.~\ref{Fig:feydg} (b)], and $u$-channel $\Sigma(1192)$ and
$\Sigma^*(1385)(\equiv\Sigma^*)$ exchanges [Fig.~\ref{Fig:feydg}
(c)].

To evaluate the invariant scattering amplitudes corresponding to
theose diagrams shown in Fig.~\ref{Fig:feydg}, the effective
Lagrangian densities for relevant interaction vertexes are needed.
Following
Refs.~\cite{Janssen:1996kx,Xie:2008ts,Doring:2010ap,Mart:2013ida,Xie:2013wfa,Xie:2013db},
the Lagrangian densities used in present work are,

\begin{eqnarray}
{\cal L}_{\pi N N} &=&-\frac{g_{\pi N N}}{2m_N} \bar{N} \gamma_5
\gamma_{\mu}\vec{\tau}\cdot \partial^{\mu}\vec{\pi} N  , \\
{\cal L}_{\pi N N^*_1} &=& -g_{\pi NN^*_1} \bar{N}^*_1 \vec{\tau} \cdot \vec{\pi} N + h.c. , \\
{\cal L}_{\pi N N^*_2} &=& -g_{\pi NN^*_2} \bar{N}^*_2 \vec{\tau} \cdot \vec{\pi} N + h.c. , \\
{\cal L}_{\pi N N^*_3} &=& \frac{g_{\pi N N^*_3}}{m_{\pi}}
\bar{N}^{* \mu}_3 \vec{\tau} \cdot \partial_{\mu} \vec{\pi} N + h.c. , \\
{\cal L}_{K\Lambda N} &=& -\frac{g_{K\Lambda
N}}{m_N+m_{\Lambda}}\bar{N} \gamma_5 \gamma_{\mu}  \partial^{\mu} K
\Lambda + h.c., \\
{\cal L}_{K\Lambda N^*_1} &=& -g_{K\Lambda N^*_1} \bar{N}^*_1 K \Lambda  + h.c., \\
{\cal L}_{K\Lambda N^*_2} &=& -g_{K\Lambda N^*_2} \bar{N}^*_2 K \Lambda  + h.c., \\
{\cal L}_{K\Lambda N^*_3} &=& \frac{g_{K \Lambda
N^*_3}}{m_{K}}\bar{N}^{*\mu}_3  \partial_{\mu} K \Lambda + h.c. ,
\end{eqnarray}
for $s$-channel nucleon pole and nucleon resonances, $N^*(1535)$,
$N^*(1650)$, $N^*(1720)$ exchange terms, and
\begin{eqnarray}
{\cal L}_{K^* \Lambda N} \! \! &=& \! \! -g_{K^*\Lambda N} \bar{N} \gamma^{\mu} K^*_{\mu} \Lambda + h.c., \\
{\cal L}_{K^*K\pi} \! \! &=& \! \! -g_{K^*K\pi} K^*_{\mu} (
\partial^{\mu}\bar{K} \vec{\tau} \! \cdot \! \vec{\pi} - \!\!
\bar{K} \vec{\tau} \! \cdot \! \partial^{\mu}\vec{\pi} ) ,
\end{eqnarray}
for the $t$-channel $K^*$ exchange process, while
\begin{eqnarray}
{\cal L}_{K N \Sigma} &=& -\frac{g_{K N \Sigma}}{m_N + m_{\Sigma}} \bar{N} \gamma_5 \gamma_{\mu} \partial^{\mu} K   \vec{\tau} \cdot \vec{\Sigma} +  h.c. , \\
{\cal L}_{\pi\Sigma\Lambda} &=& -\frac{g_{\pi\Sigma\Lambda}}{m_{\Lambda} + m_{\Sigma}} \bar{\Lambda}\gamma_5 \gamma_{\mu} \partial^{\mu} \vec{\pi}\cdot \vec{\Sigma} + h.c. , \\
{\cal L}_{KN\Sigma^*} &=& \frac{g_{K N \Sigma^*}}{m_{K}} \bar{\Sigma}^{*\mu} \partial_{\mu} K N + h.c., \\
{\cal L}_{\pi  \Lambda \Sigma^*} &=& \frac{g_{\pi \Lambda
\Sigma^*}}{m_{\pi}} \bar{\Sigma}^{*\mu}
\vec{\tau}\cdot\partial_{\mu}\vec{\pi} \Lambda + h.c. ,
\end{eqnarray}
for the $u$-channel $\Sigma(1192)$ and $\Sigma^*(1385)$ exchange
diagrams.

For the coupling constants in the above Lagrangian densities for
$t$-channel and $u$-channel processes, we take $g_{\pi NN} = 13.45$
(obtained by $g^2_{\pi NN}/4\pi = 14.4$), $g_{\rho NN} = 3.25$
(obtained by $g^2_{\rho NN}/4\pi = 0.84$), $g_{\rho \pi \pi} = 6.04$
(obtained by $g^2_{\rho \pi \pi}/4\pi = 2.90$), and $g_{\Delta N
\pi} = 2.18$~\footnote{Which is obtained by the partial decay width
of $\Delta$ to $\pi N$.}, which are used in
Ref.~\cite{Ronchen:2012eg}. The others are obtained by $SU(3)$
flavor symmetry as shown in Table~\ref{Tab:cctu}.

\begin{table}[htbp]
\caption{The coupling constants for $t$-channel and $u$-channel
processes used in the present study, which are obtained by the
$SU(3)$ flavor symmetry with $\alpha_1 = 0.4$ and $\alpha_2 =
1.15$.} \label{Tab:cctu}
\begin{tabular}{|cccccc|}
\hline
Vertex & $g$ & Value  & Vertex & $g$ & Value \\
\hline
$KN \Lambda $ & $-\frac{g_{\pi NN}}{\sqrt{3}} (1+2\alpha_1)$      & $-13.98$  & $K^* K \pi$  & $-\frac{g_{\rho \pi \pi}}{2}$    & $-3.02$  \\
$KN \Sigma $  & $g_{\pi NN}(1-2\alpha_1)$ & $2.69$  & $\Sigma^* \pi  \Lambda$ & $\frac{g_{\Delta \pi N}}{\sqrt{2}}$   & $1.54$  \\
$\pi \Lambda \Sigma$ & $\frac{2g_{\pi NN}}{\sqrt{3}}(1-\alpha_1)$    & $9.32$   & $\Sigma^* K N$ & $-\frac{g_{\Delta \pi N}}{\sqrt{6}}$ & $-0.89$ \\
$K^* \Lambda N$ & $-\frac{g_{\rho NN}}{\sqrt{3}} (1+2\alpha_2) $ & $-6.19$   &  &     &  \\
\hline
\end{tabular}
\end{table}

Besides, the coupling constants for the $s$-channel nucleon
resonance exchange processes, are obtained from the partial decay
widths,
\begin{eqnarray}
\Gamma[N^*_1 \to \pi N] &=& \frac{3 g^2_{\pi N N^*_1}}{4\pi} \frac{(E_N+m_N)}{m_{N^*_1}} |\vec{p}^N_1| , \\
\Gamma[N^*_2 \to \pi N] &=& \frac{3g^2_{\pi NN^*_2}}{4\pi} \frac{(E_N+m_N)}{m_{N^*_2}}|\vec{p}^N_2| , \\
\Gamma[N^*_2 \to K \Lambda] &=& \frac{g^2_{K \Lambda N^*_2}}{4\pi} \frac{(E_{\Lambda} + m_{\Lambda})}{m_{N^*_2}}|\vec{p}^{\Lambda}_2| , \\
\Gamma[N^*_3 \to \pi N] &=& \frac{3g^2_{\pi NN^*_3}}{12\pi} \frac{(E_N+m_N)}{m_{N^*_3}m^2_{\pi}}|\vec{p}^N_3|^3 , \\
\Gamma[N^*_3 \to K \Lambda] &=& \frac{g^2_{K \Lambda N^*_3}}{12\pi}
\frac{(E_{\Lambda}+m_{\Lambda})}{m_{N^*_3
m^2_{K}}}|\vec{p}^{\Lambda}_3|^3 ,
\end{eqnarray}
with
\begin{eqnarray}
|\vec{p}^N_i| &=& \frac{\lambda^{\frac{1}{2}}(m^2_{N^*_i},m_{\pi}^{2},m_{N}^{2})}{2m_{N^*_i}}, \\
|\vec{p}^{\Lambda}_i|
&=&\frac{\lambda^{\frac{1}{2}}(m^2_{N^*_i},m_{K}^{2},m_{\Lambda}^{2})}{2m_{N^*_i}},
\end{eqnarray}
where $\lambda$ is the K\"allen function with $\lambda(x,y,z) =
(x-y-z)^2 - 4yz$.

With masses and partial decay widthes of the nucleon resonances, the
strong coupling constants of nucleon resonances $N^*(1535)$,
$N^*(1650)$ and $N^*(1720)$ are obtained as listed in
Table~\ref{Tab:ccs}. Moreover, the strong coupling constants $g_{K
\Lambda N^*(1535)}$ is a free parameter, which will be determined by
fitting to the experimental data of the $\pi^- p \to K^0 \Lambda$
reaction.

\begin{table*}[htbp]
\caption{Relevant nucleon resonance coupling constants used in
present work. The widths and branching ratios are taken from the
Particle Data Group~\cite{pdg2014}.}\label{Tab:ccs}
\begin{tabular}{ccccccccccccc}
\hline
 Resonance ($J^p$) & ~~ Mass (GeV)~~   & ~~ Width (GeV)~~ & ~~Decay channel~~& ~~Branching ratio~~  &~~Adopted value~~ & $g^2/4\pi$\\
\hline
$ N(1535)(\frac{1}{2}^-)$ &1.535& 0.15 &$N\pi$ &  $0.35\sim0.55$ & $0.45$ & 0.037 \\
$ N(1650)(\frac{1}{2}^-)$ &1.655& 0.15 &$N\pi$ &  $0.50\sim0.90$ & $0.70$ & 0.052 \\
                          & &  &$\Lambda K$ &  $0.03\sim0.11$ & $0.07$ & 0.045 \\
$ N(1720)(\frac{3}{2}^+)$ &1.720& 0.25 &$N\pi$ &  $0.08\sim0.14$ & $0.11$ & 0.0022 \\
                          &&   &$\Lambda K$&  $0.01\sim0.15$ & $0.08$ & 0.52 \\
\hline
\end{tabular}
\end{table*}

Due to the fact that the relevant hadrons are not point-like
particles, form factors are included. In our present calculation, we
adopt the following form factors,
\begin{eqnarray}
F_{i}(q_i^2) = [\frac{\Lambda_{i}^4}{\Lambda_{i}^4+(q_{i}^2-m_{i}^2)^2}]^n ,i= s, t, u, R  \\
{\rm with} ~~\begin{cases}
q_s^2=q_R^2=s, \, q_t^2= t, \, q_u^2= u, \\
M_s = m_N, \,  M_{\rm R} = M_{N^*},\\
M_t = m_{K^*},\\
M_u = m_{\Sigma}~~{\rm or}~~m_{\Sigma^*} ,\nonumber
\end{cases}
\end{eqnarray}
with $n=1$ for $s$-channel $N^*(1535)$ and $N^*(1650)$, and $n=2$
for $s$-channel $N^*(1720)$, nucleon pole, $t$-channel $K^*$
exchange, $u$-channel $\Sigma$ hyperon pole, and $\Sigma^*(1385)$
exchange. The $s$, $t$ and $u$ are the Lorentz-invariant Mandelstam
variables, while $q_s = q_R = p_1 + p_2 = p_3 + p_4$, $q_t = p_1 -
p_3$, and $q_u = p_4 - p_1$ are the four-momenta of the intermediate
particle in the $s$, $t$, and $u$-channel, and $p_1$, $p_2$, $p_3$
and $p_4$ are the four momenta of $\pi^-$, $p$, $K^0$ and $\Lambda$,
respectively. For the cutoff parameters, they will be determined by
fitting them to the experimental data.

For propagators $G_{1/2}(p)$ of the spin-1/2 particle and
$G_{3/2}^{\mu\nu}(p)$ of the spin-3/2 particle, we adopt the simple
Breit-Wigner formula,,
\begin{eqnarray}
G_{1/2}(p) &=& i\frac{\Slash p + M}{p^2 - M^2 + iM\Gamma^{\rm {full}}} , \\
G_{3/2}^{\mu\nu}(p) &=& i\frac{(\Slash p + M)P^{\mu\nu}(p)}{p^2 -
M^2 + iM\Gamma^{\rm{full}}} ,
\end{eqnarray}
with
\begin{eqnarray}
P^{\mu\nu}(p)&=&g^{\mu\nu}-\frac{1}{3}\gamma^{\mu}\gamma^{\nu}-\frac{1}{3M}(\gamma^{\mu}p^{\nu}-\gamma^{\nu}p^{\mu}) \nonumber  \\
&& -\frac{2}{3M^2}p^{\mu}p^{\nu} ,\label{t18}
\end{eqnarray}
where $M$ and $\Gamma^{\rm{full}}$ stand for the mass and full decay
width of the corresponding resonances. It is worth to note that for
$s$-channel nucleon pole and $u$-channel $\Sigma(1192)$ and
$\Sigma^*(1385)$ exchange, we take $\Gamma^{\rm{full}} = 0$.

While for the $K^*$ meson propagator $G^{\mu\nu}_{K^*}(p)$, we take
\begin{eqnarray}
G^{\mu\nu}_{K^*}(p) &=& i\frac{-g^{\mu\nu} +
p^{\mu}p^{\nu}/m^2_{K^*}}{p^2-m^2_{K^*}} .
\end{eqnarray}

Then the total invariant scattering amplitude for $\pi^- p \to K^0
\Lambda$ reaction, according to these contributions shown in
Fig.~\ref{Fig:feydg}, can be written as,
\begin{eqnarray}
{\cal M}_{\pi^- p \to K^{0}\Lambda} &=&  {\cal M}_{\rm BG} + e^{i
\theta_1} {\cal M}_{N^*(1535)} + e^{i \theta_2} {\cal
M}_{N^*(1650)} \nonumber \\
&& + e^{i \theta_3} {\cal M}_{N^*(1720)} , \label{t34}
\end{eqnarray}
with
\begin{eqnarray}
{\cal M}_{\rm BG} &=& {\cal M}_{N} + {\cal M}_{K^*} + {\cal
M}_{\Sigma} + {\cal M}_{\Sigma^*} .
\end{eqnarray}

Note that we have employed the phase factor for the nucleon
resonances $e^{i\theta_{1,2,3}}$, since we can not determine the
phase for the nucleon resonances within our model. Thus, these phase
angles $\theta$ will be determined to reproduce the experimental
data.

With the above effective Lagrangian densities, we can
straightforwardly evaluate the following invariant scattering
amplitudes, corresponding to the Feynman diagrams in
Fig.~\ref{Fig:feydg}:
\begin{eqnarray}
{\cal M}_{N^*_{1,2}} &=& \sqrt{2} g_{\pi N N^*_{1,2}}  g_{K \Lambda
N^*_{1,2}} F_{R}(q_R^2)\bar{u}_{\Lambda}(p_4, s_{\Lambda}) \nonumber \\
&& G_{1/2}(q_R) u_{p}(p_2,s_p), \\
{\cal M}_{N^*_3} &=& -\frac{\sqrt{2} g_{\pi N N^*_3}  g_{K \Lambda
N^*_3} F_{R}(q_R^2) }{m_K m_{\pi}}
\bar{u}_{\Lambda}(p_4,s_{\Lambda}) p_{3 \mu} \nonumber
\\ && G_{3/2}^{\mu\nu}(q_R) p_{1 \nu}u_{p}(p_2,s_p), \\
{\cal M}_{N} &=& \frac{\sqrt{2} g_{\pi NN}  g_{K \Lambda N} F_{N}(q_s^2)}{2m_N(m_{\Lambda}+m_N)} \bar{u}_{\Lambda}(p_4,s_{\Lambda}) \Slash p_{3} \gamma_5  \nonumber \\
&& G_{1/2}(q_s) \gamma_5 \Slash p_{1} u_{p}(p_2,s_p) , \\
{\cal M}_{K^*} &=& -ig_{K^*\Lambda N}  g_{K^*K\pi} F_{t}(q_t^2) \bar{u}_{\Lambda}(p_4,s_{\Lambda}) \gamma_{\mu} \nonumber \\
&&  G^{\mu\nu}_{K^*}(q_t)(p_{3 \nu} + p_{1 \nu}) u_{p}(p_2,s_p), \\
\mathcal{M}_{\Sigma} &=&-g_{\pi \Lambda \Sigma}  g_{KN\Sigma} F_{u}(q_u^2) \bar{u}_{\Lambda}(p_4,s_{\Lambda}) \gamma_5  \nonumber \\
&& G_{1/2}(q_u)\gamma_5 u_{p}(p_2,s_p) , \\
{\cal M}_{\Sigma^*} &=& -\frac{g_{\pi \Lambda \Sigma^*} g_{K N
\Sigma^*}}{m_K m_{\pi}}
F_{u}(q_u^2)\bar{u}_{\Lambda}(p_4,s_{\Lambda}) p_{1 \mu} \nonumber \\
&& G_{3/2}^{\mu\nu}(q_u) p_{3 \nu} u_{p}(p_2,s_p) ,
\end{eqnarray}
where $s_{p}$ and $s_{\Lambda}$ are the polarization variables of
proton and $\Lambda$.

The unpolarized differential cross section in the center-of-mass
(c.m.) frame for the $\pi^- p \to K^0 \Lambda$ reaction reads,
\begin{equation}
\frac{d\sigma}{d\Omega} = \frac{d\sigma}{2\pi d{\rm cos}\theta} =
\frac{m_{\Lambda}m_{N}}{32\pi^{2}s}\frac{|\vec{p}^{~\rm
c.m.}_3|}{|\vec{p}^{~\rm c.m.}_1|} \left |{\cal M}_{\pi^- p \to
K^{0}\Lambda}\right|^{2} ,
\end{equation}
with $\theta$ is the polar scattering angle of outgoing $K^0$ meson,
and $\vec{p}^{~\rm c.m.}_1$ and $\vec{p}^{~\rm c.m.}_3$ are the
$\pi^-$ and $K^0$ mesons c.m. three momenta. The differential cross
section $d\sigma/d{\rm cos}\theta$ depends on $W=\sqrt{s}$ and also
on ${\rm cos}\theta$.

As mentioned above, the model accounts for a total of three
mechanisms: $s$-channel nucleon pole and $N^*$ resonances terms,
$t$-channel $K^*$ exchange, and the $u$-channel $\Sigma(1192)$ pole
and $\Sigma^*(1385)$ contributions. In principle, the free
parameters of the model are: i) relative phases between different
contributions, ii) the cut off parameters, $\Lambda_N = \Lambda_t =
\Lambda_u \equiv \Lambda_B$, and $\Lambda_{N^*(1535)} =
\Lambda_{N^*(1650)} = \Lambda_{N^*(1720)} \equiv \Lambda_R$, and
iii) the coupling constant $g_{N^*(1535)K \Lambda}$.

In the next section, we will fit the parameters of the model to the
low energy differential cross section data of the $\pi^- p \to K^0
\Lambda$ reaction by using the MINUIT fitting program.

\section{RESULTS AND DISCUSSION}

First, by including the contributions from $s$-channel nucleon pole,
$N^*(1535)$, $N^*(1650)$ and $N^*(1720)$, $t$-channel $K^*$
exchange, and $u$-channel $\Sigma(1192)$ and $\Sigma(1385)$ exchange
terms, we perform a four-parameter [$g_{K\Lambda
N^*(1535)},\theta_1,\theta_2,\theta_3$] $\chi^2$ fit (Fit I) to the
experimental data~\cite{Bertanza:1962pt,Knasel:1975rr,Baker:1978qm}
on differential cross sections of $\pi^- p \to K^0 \Lambda$
reaction. The experimental data base contains differential cross
sections at 20 energy points from 3 different experiments, and there
is a total of $385$ data points below $W = 1.76$ GeV. Second, for
showing the important role played by the $N^*(1535)$ resonance, we
have also performed another best fit, where the $s$-channel
$N^*(1535)$ resonance has been switched off (Fit II).

The fitted parameters of Fit I and Fit II are compiled in Table
~\ref{Tab:fitpara}. The resultant $\chi^2/dof$ of Fit I is $2.3$,
while for the Fit II, it is $3.4$, which is turn out to be larger,
since we have ignored the important contributions from the
$N^*(1535)$ resonance. From the the fitted result, $g_{K\Lambda
N^*(1535)} = 1.31 \pm 0.08$, we obtain the ratio $R \equiv | g_{K
\Lambda N^*(1535)}/g_{\eta N N^*(1535)} | = 0.71 \pm 0.10$ with the
value of $g_{\eta N N^*(1535)} = 1.85 \pm 0.24$ that is obtained
with the partial decay width of $N^*(1535) \to N\eta$. This value is
smaller than the one $1.3 \pm 0.3$ obtained in
Ref.~\cite{Liu:2005pm} by analyzing the $J/\psi \to \bar{p}K^+
\Lambda$ and $J/\psi \to \bar{p}\eta p$ experimental data. This is
because in the work of Ref.~\cite{Liu:2005pm}, the energy dependent
width for the $N^*(1535)$ resonance is used, which will decrease the
contribution from the propagator of $N^*(1535)$ resonance above the
$K\Lambda$ threshold and make the coupling strength of the
$N^*(1535)$ resonance to the $K \Lambda$ channel
larger.~\footnote{Similar statement can be also found in
Ref.~\cite{Xie:2013wfa} for the case of the $\Lambda(1405)$.} In
contrast with the energy dependent width as in
Ref.~\cite{Liu:2005pm}, in the present work we use a constant total
decay width for $N^*(1535)$ resonance since the $K \Lambda$ channel
is opened. Nevertheless, the values of this ratio obtained in the
previous works are widely scattered. For instance, $R = 0.460 \pm
0.172$ has been determined from the latest and largest
photoproduction database by using the isobar model. Similar results,
$R= 0.5 \sim 0.7$ were obtained in Ref.~\cite{Geng:2008cv} from the
$J/\psi$ decays within the chiral unitary approach and $R= 0.42 \sim
0.73$ was obtained in Ref.~\cite{Sarantsev:2005tg} by the partial
wave analysis of kaon photonproduction. In Ref.~\cite{Bruns:2010sv}
the result of the $s$-wave $\pi N$ scattering analysis within a
unitarized chiral effective Lagrangian indicates that $|g_{ g_{K
\Lambda N^*(1535)}}|^2 > |g_{\eta N N^*(1535)}|^2$, whereas a
coupled-channels calculation predicted a value of $R = 0.8 \sim
2.6$~\cite{Penner:2002md}.

\begin{table}[htbp]
\begin{center}
\caption{Fitted parameters of Fit I and Fit II.}\label{Tab:fitpara}
\begin{tabular}{|ccc|}
\hline
Parameters & Fit I & Fit II \\
\hline
$\Lambda_B$ [GeV]    & $0.86 \pm 0.02$  & $0.80 \pm 0.01 $ \\
$\Lambda_R$ [GeV]    & $0.85 \pm 0.04$  & $0.78 \pm 0.03 $ \\
$g_{K\Lambda N^*(1535)}$  & $1.31 \pm 0.08 $ & -- \\
$\theta_1$      & $2.59 \pm 0.08 $ &-- \\
$\theta_2$      & $1.40 \pm 0.08 $ & $ 0.60 \pm 0.06 $ \\
$\theta_3$      & $0.26 \pm 0.07 $ & $-0.94 \pm 0.05 $ \\
$\chi^2/dof$    & $2.3$            & $3.4$             \\
\hline
\end{tabular}
\end{center}
\end{table}

\begin{figure*}[htbp]
\begin{center}
\includegraphics[scale=0.45]{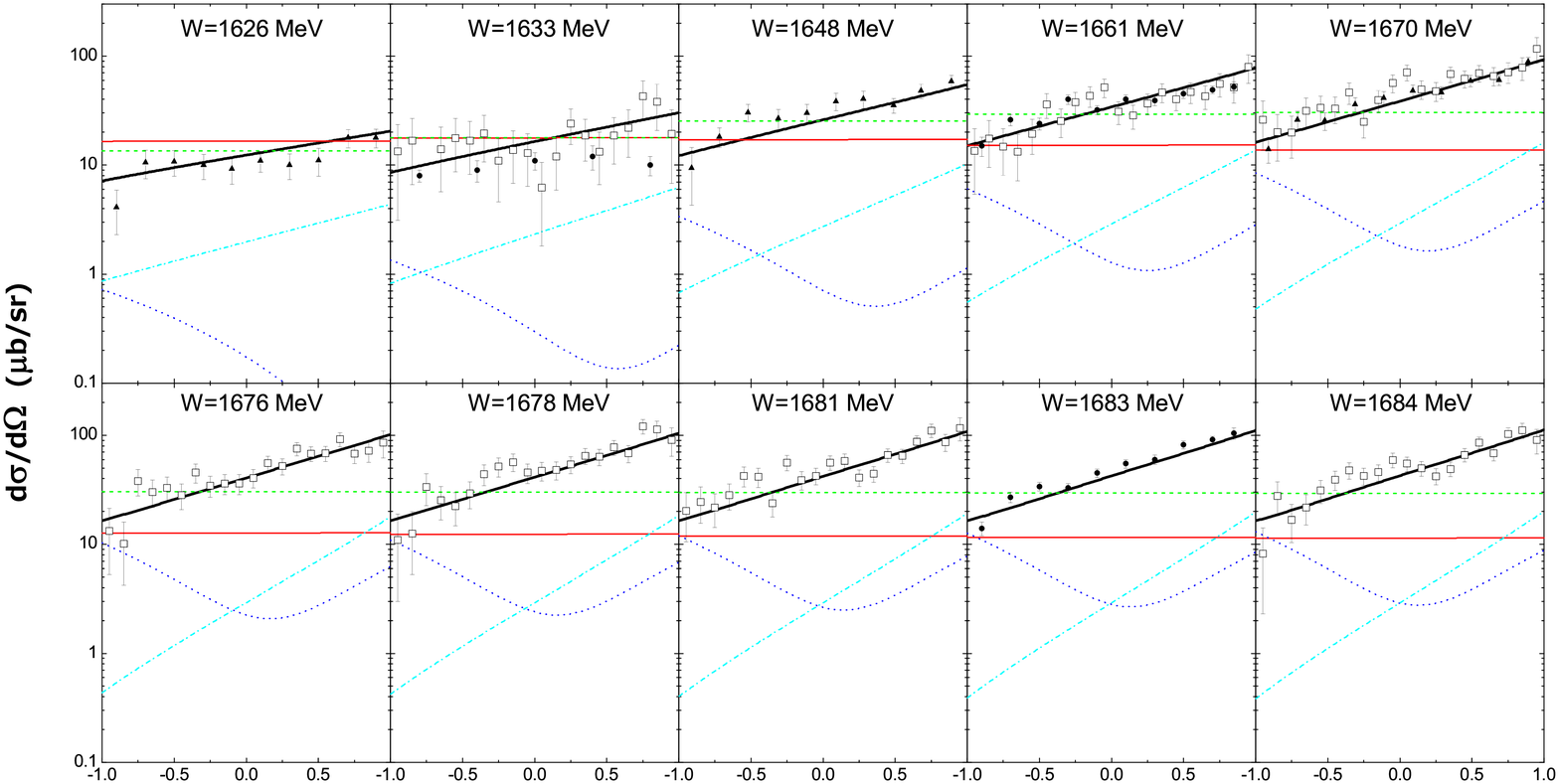}
\includegraphics[scale=0.45]{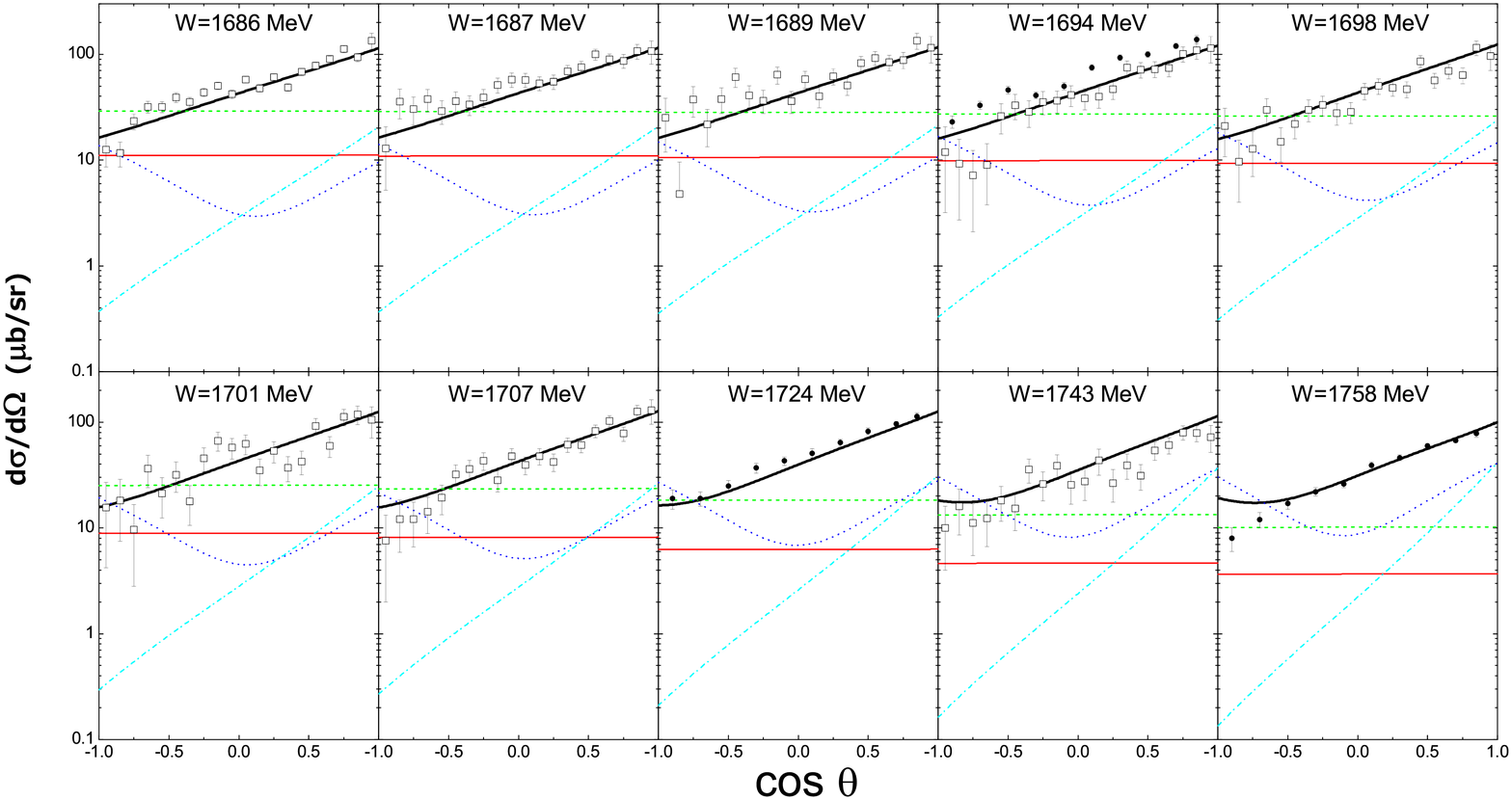}
\caption{(Color online) Fit I differential cross sections for the
reaction of $\pi^- p \to K^{0}\Lambda$ in the center of mass frame
as a function of ${\rm cos}\theta$ at different $\pi^- p$ invariant
mass intervals compared with the experimental data from
Ref.~\cite{Knasel:1975rr} (squares), Ref.~\cite{Baker:1978qm}
(circles), Ref.~\cite{Bertanza:1962pt} (triangles).  The cyan
dashed-dotted, red solid, green dashed, and blue dotted curves stand
for the contributions from the the background ($s$-channel nucleon
pole, $t$-channel $K^*$ exchange and $u$-channel $\Sigma$ and
$\Sigma^*(1385)$ exchange terms), $s$-channel $N^*(1535)$,
$N^*(1650)$, and $N^*(1720)$ terms, respectively. The bold black
solid lines represent the results obtained from the full model.}
\label{Fig:dcs}
\end{center}
\end{figure*}

The differential distributions $d\sigma/d\Omega$ calculated with the
Fit I best-fit parameters are shown in Fig.~\ref{Fig:dcs} as a
function of ${\rm cos}\theta$ and for various $\pi^- p$ invariant
mass intervals. The contributions from different mechanisms are
shown separately. Thus, we split the full result into four main
contributions: effective Lagrangian approach background, $s$-channel
$N^*(1535)$, $N^*(1650)$ and $N^*(1720)$. The first one corresponds
to the $t$-channel $K^*$ exchange, $s$-channel nucleon pole, and
$u$-channel $\Sigma(1192)$ hyperon pole and $\Sigma^*(1385)$ terms.
We find an overall good description of the data for the whole range
of measured $\pi^- p$ invariant masses below $1.76$ GeV, and the
contributions from the above three main mechanisms are all
significant to reproduce the current experimental data.

\begin{figure}[htbp]
\begin{center}
\includegraphics[scale=0.2]{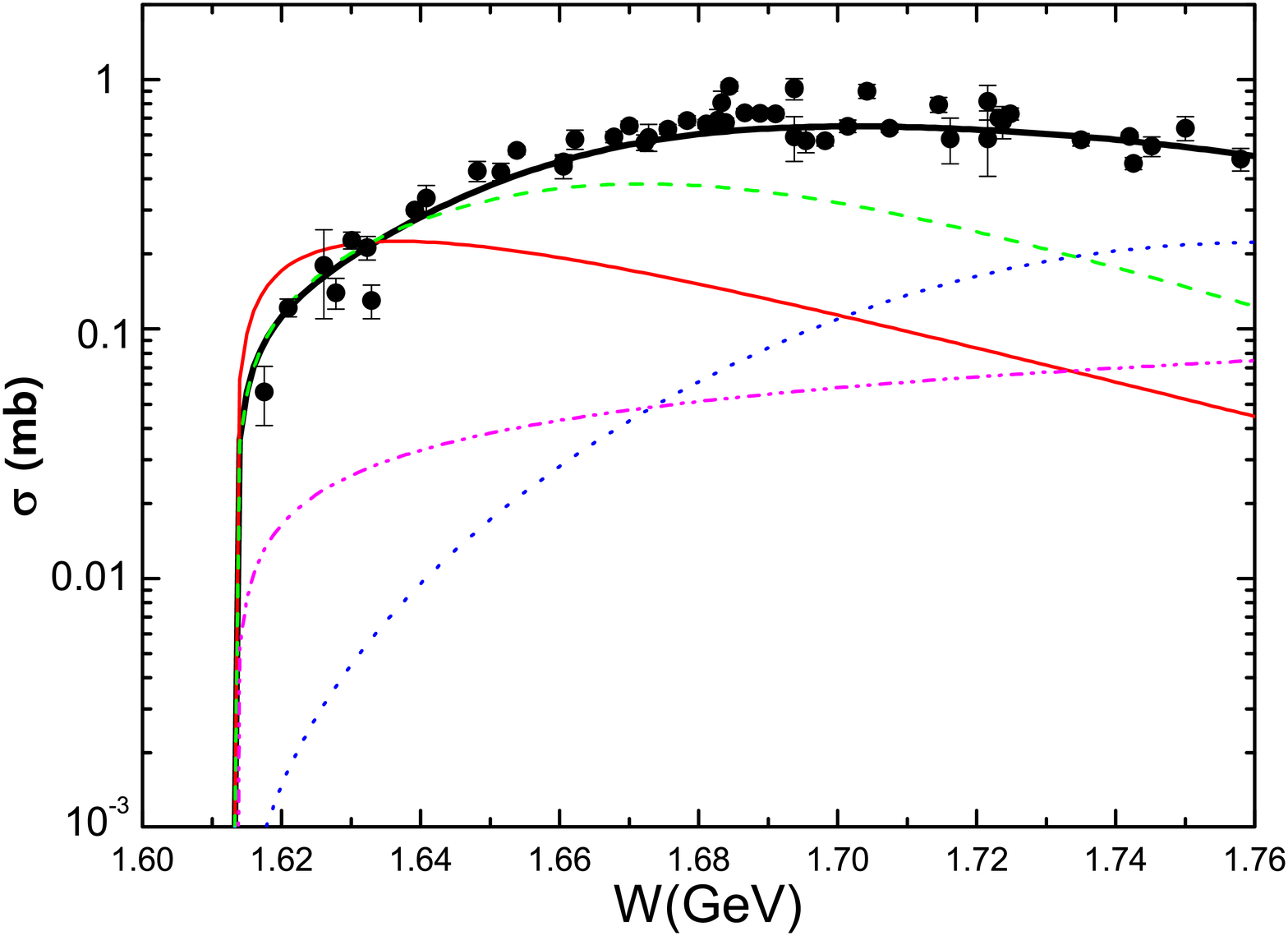}
\caption{(Color online) Total cross section of the $\pi^- p \to
K^{0}\Lambda$ reaction as a function of the invariant mass of $\pi^-
p$ system with the fitted parameters of Fit I. The experimental data
are taken from Ref.~\cite{Baldini:1988ti}. The bold black solid
lines represent the results from the full model, while the
contributions from $t$-channel $K^*$ exchange and $s$-channel
$N^*(1535)$, $N^*(1650)$, $N^*(1720)$ terms are shown by the magenta
dash-dot-doted, red solid, green dashed and blue doted curves,
respectively.} \label{Fig:tcs}
\end{center}
\end{figure}

Our best result of the total cross sections of $\pi^- p \to K^0
\Lambda$ reaction as a function of the invariant mass of the $\pi^-
p$ system are shown in Fig.~\ref{Fig:tcs} compared with the
experimental data~\cite{Baldini:1988ti}. There, we see that we can
describe the total cross section data quite well. As mentioned
above, the $N^*(1535)$ resonance and $N^*(1650)$ resonance couple
strongly to the $K\Lambda$ channel. Indeed, the total cross section
of $\pi^- p \to K^0 \Lambda$ show a strong $S$-wave contributions
close to the reaction threshold and also a little bit beyond. From
Fig.~\ref{Fig:tcs}, it is seen that the contribution from
$N^*(1650)$ (blue dashed curve) is predominant at a very wide energy
region, and the contribution from the $N^*(1535)$ (red solid curve)
is significant from the reaction threshold till $W < 1.66$ GeV,
while the contributions from $t$-channel $K^*$ exchange and
$s$-channel $N^*(1720)$ are dominant above $W > 1.7$ GeV. The
$s$-channel nucleon pole and $u$-channel contributions are too small
to be shown in Fig.~\ref{Fig:tcs} and can be neglected.

It is worthy to note that we do not consider the contribution from
the nucleon resonance $N^*(1710)$, which is mostly required by the
$\pi N$ inelastic scattering data. The role of this resonance in the
context of different partial-wave analyses has been discussed
extensively in Refs.~\cite{Ronchen:2012eg,Ceci:2006ra}. In the
analysis of the $\pi^- p \to n \eta$ reaction, a $N^*(1710)$
resonance is needed~\cite{Ceci:2006ra}. But, it is pointed out in
Ref.~\cite{Ronchen:2012eg} that in the analysis of the $\pi^- p \to
K^0 \Lambda$ reaction, there is an interplay between the $N^*(1720)$
and $N^*(1710)$ and individual contributions are difficult to pin
down. Moreover, the $N^*(1710)$ resonance appears in the three-body
hadronic
calculations~\cite{Khemchandani:2008rk,MartinezTorres:2008kh} and
also could be a dynamically generated
resonance~\cite{Suzuki:2009nj,Kamano:2010ud}. On the other hand, a
Bayesian analysis of the world data on $\gamma p \to K^+ \Lambda$
reaction~\cite{DeCruz:2011xi} show that there is no significant
contribution of the $N^*(1710)$ resonance to the $\gamma p \to K^+
\Lambda$ reaction. Because of those doubts of this resonance, we
ignore its contribution in our present calculations.

\section{SUMMARY}

In this paper, the $\pi^- p \to K^{0}\Lambda$ reaction is
investigated within an effective Lagrangian approach and the isobar
model. This channel is known to receive significant nonresonant
contributions which complicates the extraction of $N^*$ information.
In addition to the background contributions from the $s$-channel
nucleon pole, $t$-channel $K^*$ exchange, $u$-channel $\Sigma(1192)$
and $\Sigma^*(1385)$ exchanges, we also consider the contributions
from $s$-channel nucleon resonances $N^*(1535)$, $N^*(1650)$, and
$N^*(1720)$. From $\chi^2-$fits to the available experimental data
for the $\pi^- p \to K^0 \Lambda$ reaction, we get the appropriate
parameters which describe the total and differential cross sections
well. Our results show that the inclusion of the nucleon resonances
$N^*(1535)$, $N^*(1650)$, and $N^*(1720)$ can lead to a good
description of the low energy experimental total and differential
cross sections data of $\pi^- p \to K^0 \Lambda$ reaction. The
contribution from each individual resonance to the total and
differential cross sections below $\sqrt{s}=1.76$ GeV are also
presented. The contributions from those nucleon resonances and the
$t$-channel $K^*$ exchange are dominant, while $s$-channel nucleon
pole and the $u$-channel $\Sigma(1192)$ and $\Sigma^*(1385)$
exchange give the minor contributions and can be neglected.

\section*{Acknowledgments}

We would like to thank Xu Cao for useful discussions. This work is
partly supported by the Ministry of Science and Technology of China
(2014CB845406), the National Natural Science Foundation of China
under grants: 11105126, 11375024 and 11175220. We acknowledge the
one Hundred Person Project of Chinese Academy of Science
(Y101020BR0). The Project is sponsored by the Scientific Research
Foundation for the Returned Overseas Chinese Scholars, State
Education Ministry.

\end{document}